# Atomic-scale investigation of creep behavior in nanocrystalline Mg and Mg-Y alloys

M. A. Bhatia[1], S.N. Mathaudhu[2], and K.N. Solanki[1*]

[1]*School for Engineering of Matter, Transport, and Energy, Arizona State University, Tempe, AZ*
[2]*Materials Science and Engineering Program and Mechanical Engineering Department, University of California - Riverside, Riverside, CA*
*(480)965-1869; (480)727-9321 (fax), E-mail: kiran.solanki@asu.edu, (Corresponding author)*

**Abstract**

Magnesium (Mg) and its alloys offer great potential for reducing vehicular mass and energy consumption due to their inherently low densities. Historically, widespread applicability has been limited by low strength properties compared to other structural Al-, Ti- and Fe-based alloys. However, recent studies have demonstrated high-specific-strength in a number of nanocrystalline Mg-alloys. Even so, applications of these alloys would be restricted to low-temperature automotive components due to microstructural instability under high temperature creep loading. Hence, this work aims to gain a better understanding of creep and associated deformation behavior of columnar nanocrystalline Mg and Mg-yttrium (Y) (up to 3at.%Y(10wt.%Y)) with a grain size of 5 nm and 10 nm. Using molecular dynamics (MD) simulations, nanocrystalline magnesium with and without local concentrations of yttrium is subjected to constant-stress loading ranging from 0 to 500 MPa at different initial temperatures ranging from 473 to 723 K. In pure Mg, the analyses of the diffusion coefficient and energy barrier reveal that at lower temperatures (i.e., T < ~423K) the contribution of grain boundary diffusion to the overall creep deformation is stronger that the contribution of lattice diffusion. However, at higher temperatures (T > ~423K) lattice diffusion dominates the overall creep behavior. Interestingly, for the first time, we have shown that the $(10\bar{1}1), (10\bar{1}2), (10\bar{1}3)$ and $(10\bar{1}6)$ boundary sliding energy is reduced with the addition of yttrium. This softening effect in the presence of yttrium suggests that the experimentally observed high temperature beneficial effects of yttrium addition is likely to be attributed to some combination of other reported creep strengthening mechanisms or phenomena such as formation of stable yttrium oxides at grain boundaries or increased forest dislocation-based hardening.

# I. Introduction

Increased global demand for energy-efficient yet safer materials in the automotive and aviation industries requires the development of new lightweight alloys. Among structural materials, Magnesium (Mg) has the highest strength-to-weight ratio. Moreover, Mg-alloys have excellent recyclability and are ideal for transportation applications providing enormous potential for energy savings [1–7]. However, magnesium has certain limitations when compared with other metallic systems and polymers [6]. Since magnesium has a hexagonal close packed (hcp) structure, it has a limited number of active slip systems at room temperature [8,9] which leads to failure in polycrystalline Mg-alloys at ambient conditions before dislocation glide can occur [10–12]. In an effort to overcome these low strength shortcomings, recent studies have considered the effects of adding certain alloying elements to magnesium, such as Al-Zn, Al-Si, Al-Ca, Al-Sr, Sn-Ca, Zr-Y, and minute traces of other rare-earth (RE) elements [6,13–23]. The objective of all these studies has been to qualitatively identify the principles governing solute and precipitate strengthening and ductility while improving manufacturability, as well as lowering production costs of the alloys (e.g., [17,24]). The studies have also shown that the alloying elements can cause Mg-alloys to either harden or soften by interacting with dislocations/twins/interfaces and/or forming favorable phases [24]. Rare earth elements (RE), in particular, have been found to generate the most favorable effects on the mechanical and physical properties of Mg-alloys, including improved creep and corrosion resistance [13,16–18,24–27]. The addition of RE to Mg-alloys, however, has been found to be cost prohibitive, so other solutions, such as RE reduction or replacement, are needed.

Another approach to strengthening Mg alloys has been through the route of grain refinement to the nanoscale. Recent studies have revealed extremely high hardness and strengths in excess of 500MPa in nanocrystalline Mg alloys [28–31]; however, in-depth studies of the governing high temperature deformation mechanisms and behavior in these nanocrystalline Mg-alloys (and other hcp alloys for that matter) have yet to be reported. In order to develop additional, cost-effective, high-strength alloys with suitable high temperature properties, a better understanding of the deformation mechanisms governing nanostructured Mg-alloy performance is necessary.

High-temperature deformation mechanisms, such as creep in coarse-grained magnesium and its alloys, have been extensively studied over the years, and detailed reviews can be found in [2,4,18,24]. These studies reveal clearly that higher temperature creep loading leads to reduction in the load-bearing capacity of coarse-grained Mg-alloys. Specifically, during creep, the strain is accumulated either by diffusional flow through the interfaces (Nabarro-Herring creep [32,33]), diffusional flow along interfaces (Coble creep [34]), interface sliding (Raj and Ashby [35]), or dislocation motion (climb/glide [36]). In general, these creep mechanisms/models have been vigorously debated (e.g., see [24]) mostly due to the inability of certain models to accurately predict creep behavior across materials with varying microstructures. In recent experimental studies, the important role played by microstructural attributes, such as grain size and solutes/phases segregated to the grain boundary (GB) in the creep behavior of Mg-alloys [15,18], was demonstrated. It was found that with increasing grain size, the creep rate decreased with an attendant loss in room temperature yield strength as predicted by the Hall-Petch equation [37–40]. On the other hand, grain-size refinement could be controlled by the addition of micro-alloys such as zirconium [15,41]. Thus, with a balance between goals for optimal grain size and room-temperature strength, creep deformation behavior potentially could be enhanced by elucidating the deformation mechanisms necessary to engineer GBs with solutes such as yttrium (Y).

In the case of coarse-grained Mg-Y alloys, several experimental and modeling techniques have been employed to reveal the fundamental mechanisms related to yttrium solute interaction with material microstructure under various stress-states and environments (i.e., creep, dynamic loading, etc.) as well as the mechanisms' subsequent effect on engineering properties. It has been reported that the addition of yttrium leads to *(a)* improved ductility and homogeneous deformation due to activation of <c+a> dislocations [11,20,21,42], *(b)* improved creep resistances due to forest dislocation hardening [18], *(c)* retardation of the kinetics associated with static recrystallization and grain coarsening, which suggest pinning of grain boundaries [14], *(d)* grain boundary rotation/sliding preventing the nucleation and propagation of cracks [19], and *(e)* improved corrosion resistance due to the formation of an enhanced oxide film [43,44].

In this paper, extensive molecular statics (MS) and molecular dynamics (MD) simulation results are presented to elucidate the effect of grain size (5 nm to 15 nm) and grain boundaries with various yttrium concentrations (1 to 3 at.% Y or 3.56 to 10.16 wt.% Y) on diffusional creep and deformation mechanisms of nanocrystalline magnesium. Further, to overcome the relatively short time interval of MD simulation, we also performed simulations at elevated temperatures where the distinct effects of GB diffusion (liquid-like fast GB diffusion) are clearly identifiable (see the work of [45,46]). Other notable atomistic work on the kinetic and thermodynamic minimization of grain growth includes those by Schuh, Millett, and other researchers [45–52]. This paper is organized as follows. The next section describes the computational methods employed (interatomic potentials, creep deformation modeling, and grain boundary sliding). Then, we present several different characterizations of nanocrystalline magnesium and magnesium with yttrium creep deformation behavior: (1) diffusional analysis with varying grain sizes at 0 MPa, (2) diffusional analysis with varying concentrations of Yttrium at 0 MPa, (3) diffusional analysis with varying grain sizes at 500 MPa, (4) diffusional analysis with varying concentrations of Yttrium at 500 MPa, and (5) the effect of yttrium on grain boundary sliding. First part of the paper presents an extensive study on the creep behavior of nanocrystalline magnesium at different temperatures and grain sizes. We show that for nanocrystalline magnesium samples, the analyses of the diffusion coefficient and energy barrier reveals that at lower temperatures (i.e. T < 423K) the contribution of grain boundary diffusion to the overall creep deformation is stronger that the contribution of lattice diffusion. However, at higher temperatures (T > 423K) lattice diffusion dominates the overall creep behavior. Further, we used pure nanocrystalline magnesium creep data to effectively compare creep responses of nanocrystalline Mg-Y. We observed a negligible change (within the fitting error) in the overall secondary creep rate, with creep activation energy changing from 1.128 to 1.154 eV for 0 to 3 at.% Y, respectively, indicating that stage two creep activity is insensitive to Y for a given grain size. We have studied and shown that $(10\bar{1}1)$, $(10\bar{1}2)$, $(10\bar{1}3)$ and $(10\bar{1}6)$ boundary sliding energy is reduced with the addition of yttrium. This softening effect in the presence of yttrium suggests that the experimentally observed high temperature beneficial effects of yttrium addition could be attributed to other mechanisms or phenomena such as formation of stable yttrium oxides at grain boundaries or increased forest dislocation-based hardening.

## II. Computational methods
### A. Interatomic potential and crystal properties

MS and MD calculations were performed using a large-scale atomic/molecular massively parallel simulator, LAMMPS [53]. A semi-empirical embedded atom potential (EAM), developed by Pei et al. [42], was used to describe the Mg-Y system. The interatomic potential parameters were derived from a fitting method using isothermal compressibility data up to 2000 K (1727 ˚C).

Moreover, the potential was developed by also fitting the vacancy migration energy as well as with different Mg-Y crystal structures. More details on the validation of EAM potential at different temperatures can be found in Pei et al. [42]. Table 1 lists the cohesive energy of magnesium and yttrium as well as other bulk properties and its comparison with density functional theory (DFT) data [54], our own DFT data (see the supplemental section) , and experimental results [55]. This potential has also been used to study yttrium effects on the reduction of the unstable stacking fault energy of basal slip systems and has shown good agreement with DFT and experimental results [42]. These prior findings with the aforementioned potential rationalize its application for studying the yttrium effect on the deformation of nanocrystalline Mg.

## B. Creep deformation of nanocrystalline magnesium

A parallel molecular dynamics code (LAMMPS) [53] was used to deform the nanocrystalline magnesium samples of 5-15 nm grain dimension as shown in Figure 1. The Voronoi tessellation method [56] was used to construct [11$\bar{2}$0] columnar samples of 5.6 nm thickness containing 4 grains of identical hexagonal shape within periodic simulation cells (see more detailed in [45,46]). After generating the desired magnesium nanocrystalline model, grain boundary atoms were selected based on a cohesive energy of the atom greater than -1.45 eV. Then, yttrium atoms were randomly doped along the grain boundary to obtain the desired concentrations (1~3at. % or 3~10wt. %). Doping at the boundaries was based on the McLean segregation model, which predicts the majority of Y will preferentially segregate to grain boundaries at elevated temperatures (see work of [57]).The total number of atoms in a simulation cell was in the range of 250000-1 million atoms (with approximate box sizes in the range of 30-65 nm X 35-70 nm X 5.5 nm). The samples were first relaxed at the desired temperature using an NVT (conserving the number of atoms, volume, and temperature) ensemble for 15 ns followed by an independent relaxation in three directions using an NPT (conserving the number of atoms, pressure, and temperature to mimic bulk behavior) ensemble for another 10 ns with zero pressure in all the directions. These relaxations are performed to uniformly distribute the excess free energy through the whole system. Atomistic simulations were carried out using an MD time step of 1 fs. A periodic boundary condition was adopted in all directions. A constant stress ranging from 0 to 500 MPa was applied along the Y direction at different temperatures ranging from 423 K to 723 K (150 ˚C – 450 ˚C), whereas deformation in the other two directions was carried out by maintaining a zero pressure, according to the Parrinello-Rahman method [58]. The desired stresses were applied in the incremental form (a 5 MPa step) and then simulations were run at a constant applied stress for 15 ns or until failure.

### C. Grain boundary sliding

Understanding the structure and energy of grain boundaries in engineering materials is crucial because grain boundary properties can vary widely (e.g., coherent twin versus low angle versus high angle grain boundaries). To investigate grain boundary sliding behavior, we employed MS simulations. The simulation cell used for quantifying the grain boundary sliding behavior is constructed as follows. The initial single crystals were created with x, y, and z along the $[0\bar{1}10]$, $[0001]$, and $[1\bar{2}10]$ directions, respectively, for the $[1\bar{2}10]$ tilt axis. Then the equilibrium 0 K grain boundary structure and energy was calculated using a bicrystal computational cell with three-dimensional (3D) periodic boundary conditions consisting of two grains. The minimum distance between the two periodic boundaries in each computational cell was 12 nm along the x direction. The minimum dimensions for the entire bicrystal at equilibrium were approximately 12 nm × 10 nm × 5 nm. As with past work [59–67] and others [68–70], an atom deletion criterion, multiple initial configurations, and various in-plane rigid body translations were utilized to accurately obtain an optimal minimum energy GB structure via the nonlinear conjugate gradient energy minimization process. Further details on the GB generation technique and GB structures of HCP materials are given in Bhatia and Solanki [59], and Wang and Beyerlein [71]. Figure 2a shows optimum grain boundary energies as a function of the misorientation angle for the $[1\bar{2}10]$ tilt axis (after many (100's of) different in-plane rigid body translations). The energy cusps for the $[1\bar{2}10]$ system were identified as $(10\bar{1}3)$ θ = 32.15°, $(10\bar{1}2)$ θ = 43.31°, and $(10\bar{1}1)$ θ = 62.06° twin boundaries, and $(20\bar{2}1)$ θ = 75.21° grain boundary, in order of increasing misorientation angle.

These stable (equilibrium) twin boundaries along with high angle ($(20\bar{2}1)$ θ = 75.21°) and low angle ($(10\bar{1}6)$ θ = 17.01°) grain boundaries were then selected for further analyses, and represent a smaller range of boundaries than that experimentally observed in poly/nanocrystalline magnesium. To simulate the grain boundary sliding and illustrate the effect of yttrium (grain boundaries where doped randomly at high segregation potency sites), periodic boundaries were maintained along the x and z directions; whereas, a shrink-wrapped boundary condition was prescribed along the third direction (y axis). An incremental shear displacement of 0.01% was applied on the top grain in the z ($[1\bar{2}10]$) direction. Each displacement increment was followed by a nonlinear conjugate gradient energy minimization process with force criteria of 1 pN on each atom in the top grain along the y direction. Figure 2b depicts a sketch of twin boundary sliding. The excess energy was then plotted as a function of applied shear displacement. The excess energy ($E_{ex}$) is calculated based on

$$E_{ex} = \frac{E_{dis} - E_{gb}}{A} \qquad (1)$$

where $E_{ex}$ is the excess energy required to shear the boundary, $E_{dis}$ is the total energy when the boundary is displaced by some finite amount, $E_{gb}$ is the boundary energy and A is the area of the boundary.

## III Results and Discussion
### A. Diffusional analysis with varying grain sizes at 0 MPa

To study the diffusional mechanisms of nanocrystalline Mg, equilibrated structures with different grain sizes were annealed for 500 ps in the temperature range of 473 K to 723 K under zero applied load condition. These simulations were used to calculate the mean square displacement (MSD) values as

$$MSD = \sum_{i=1}^{N}(r_i(t) - r_i(0))^2 \qquad (2)$$

at different constant temperatures, where N is the number of atoms, $r_i(t)$ is the position of an atom at any instant t, and $r_i(0)$ is the initial position of an atom at $t = 0$. For the GB simulations, we used $\pm 8$Å as an effective thickness of the GB region to calculate the GB self-diffusivity using the MSD technique. From the MSD values, the diffusion coefficient was computed with the help of Einstein's equation as follows:

$$MSD = 6Dt, \qquad (3)$$

where D is the diffusion coefficient, and *t* is the time. Diffusion in an HCP crystal is anisotropic and depends on the crystal c/a ratio as shown in the work of Mantina et al. [72]. In this study, our objective was to clarify the effect of yttrium on grain boundary diffusion and the mechanism of grain boundary deformation. Hence, the lattice diffusion of magnesium in the grain and along the grain boundary is assumed to be isotropic. Moreover, it is also understood that as each grain is oriented at a different angle to the loading orientation, the overall diffusion will be isotropic. Initially, to differentiate different dominant mechanisms (i.e., lattice versus grain boundary), we calculated the activation energy for the lattice and grain boundary (GB) diffusions; see the supplemental section for more detail. The activation energy for the bulk was found to be 0.73 eV and for the grain boundary (bi-crystal) was found to be ~0.55 eV (see the supplemental section).

Figure 3 shows Arrhenius plots for self-diffusion of all magnesium atoms (including the grain boundary atoms contributions) in nanocrystalline magnesium with different grain sizes that are given as follows:

$$D = D_o e^{-Q/kT} \qquad (4)$$

where $D_o$ is the maximum diffusion coefficient at infinite temperature, $Q$ is the activation energy, $k$ is the Boltzmann constant, and $T$ is the temperature. In order to obtain values of absolute low temperature diffusion and high temperature diffusion coefficients, activation penetration curves were obtained at various temperatures. To compute two distinct slopes, we used Johnson's relationship [73] along with Fisher's analysis [74], i.e., the logarithmic activities varies linearly with the depth of penetration rather than with quadratic. For a lower temperature range, the slope of whole system diffusion is closer to the grain boundary atom diffusion slope (see the supplemental section). For example, the activation energy for the 15 nm columnar grain model was 0.45 eV, which is comparable to the 0.55 eV activation energy found using bi-crystal models. The difference in activation energies at a lower temperature range is because the diffusion contribution from a fraction of grain boundary atoms (0.020) in a columnar grain is greater than the fraction of grain boundary atoms (0.014) in the bi-crystal model. In contrast, for a higher temperature range, the slope is closer to the bulk atom slope. The activation energy for the bulk was found to be 0.73 eV (see the supplemental section), which is comparable with the activation energy of 0.714 eV that was found for a 10 nm columnar grain model at high temperature (Figure 3). Hence, at lower temperature, grain boundary atoms dominate the diffusion process, whereas at higher temperature (> 0.7$T_m$), bulk atoms contribute significantly to the diffusion process. For the remainder of the manuscript, the high temperature diffusion is labeled as "lattice diffusion" and the low temperature diffusion is labeled as "grain boundary diffusion".

As the grain size increases, the activation energy for GB diffusion and for lattice diffusion also increases. Table 2 lists the respective activation energies for all magnesium atoms (including the grain boundary atoms contributions) in nanocrystalline magnesium with different grain sizes. It is to be noted that the fraction of GB atoms increases as the grain size decreases. Hence, with smaller grain sizes, the contribution from the GB diffusion increases and dominates the creep deformation mechanism as predicted by the activation energy ratio $Q_R = Q_{lt}/Q_{ht}$, where $Q_{lt}$ is the diffusion activation energy of the whole system at lower temperature, and $Q_{ht}$ is the diffusion activation energy of the whole system at higher temperature. For the nanocrystalline magnesium with 5 nm grain size, $Q_R$ is equal to 0.558, and hence, the activation energy of low temperature diffusion (0.384 eV) is much lower than the activation energy of high temperature diffusion (0.688 eV). As the grain size increases, the activation energy also increases, but the rate of increase in activation energy for the high temperature diffusion is slower than for the low temperature diffusion. Moreover, the fraction of the contribution of the GB atom to total diffusion also decreases. Hence, at 15 nm grain size, $Q_R$ is equal to 0.599. Further work is required to find out the grain size at which there is a transition from the GB diffusion to the lattice diffusion mechanism.

## B. Diffusional analysis with varying concentrations of Yttrium at 0 MPa

To monitor the diffusion paths of the GB atoms, the equilibrated 5 nm grain nanocrystalline structure with different initial concentration of yttrium was annealed for 250 ps under zero applied stress in the temperature range of 423 K to 723 K. These equilibrated structures were used to calculate mean square displacement (MSD) at given temperatures of interest. Figure 4 shows the MSD evolutions for the magnesium and yttrium atoms at grain boundaries for different temperatures. These results suggest a linear increase in MSD values for both magnesium and yttrium, with magnesium being more mobile than yttrium.

Figure 5 shows the Arrhenius plots for diffusion of magnesium and yttrium at the grain boundary in 5nm grain size Mg-1at.%Y (or 3.5wt.%Y). Figure 3 is related to diffusion of the total system (including grain boundaries); whereas, Figure 5 is diffusion of just grain boundary atoms. In the case of Mg-1at.%Y (or 3.5wt.%Y) in a low temperature range, the activation energy for diffusion of yttrium, 0.094 eV, is lower than that of magnesium, 0.279 eV. In contrast, for a higher temperature range, the activation energy for diffusion of yttrium, 0.93 eV, is higher than that of magnesium, 0.659 eV. However, in the case of Mg-3at.%Y (or 10.16wt.%Y) at high temperature, the activation energy for diffusion of magnesium and yttrium decreases when compared with Mg-1at.%Y (or 3.56wt.%Y). Table 3 lists the respective activation energies for diffusion of magnesium and yttrium at the grain boundary in nanocrystalline magnesium with different yttrium concentrations. From table 3, we can conclude that the addition of yttrium increases the activation energy for grain boundary magnesium atoms compared to the activation energy of grain boundary magnesium atoms with no yttrium. Hence, the mobility of magnesium decreases with the addition of yttrium. This has general implications on the grain boundary decohesion and grain boundary sliding energy which will be discussed later in the diffusional and grain boundary sections.

### C. Diffusional analysis with varying grain sizes at 500 MPa

To study the effect of applied loads, a constant applied stress of 500 MPa was applied in the y direction, and lateral dimensions were held at 0 MPa. The magnitude of the applied stress was low enough to ensure that no dislocations were nucleated from the GBs. Figure 6 shows the strain as a function of time at 623 K with different grain sizes. Creep behavior in Figure 6 is divided into three regions, as is observed in many materials. In Region I (primary creep), there is an instantaneous response, followed by a decreasing rate of strain with an increase in time. In Region II (secondary creep), the strain increases linearly with time. In Region III (tertiary creep), the strain increases rapidly with time. Most of the damage or fracture is experimentally observed in Region III due to voids and crack nucleation. We also observe, as shown in Figure 6, that the secondary creep in the 5 nm grain sample is short-lived in comparison to the 10 nm and 15 nm grain samples.

Based on these results, the 10 nm grain sample was used to explore further the effect of yttrium on the steady state creep regime. The main objective here was to determine whether alloying elements such as yttrium could delay tertiary creep by prolonging the secondary creep regime.

An empirical equation, which describes the strain-time relation, is given by

$$\dot{\varepsilon} = C \left(\frac{\sigma}{G}\right)^m \frac{1}{d^b} e^{-Q/kT} \qquad (5)$$

where $m$ and $b$ are the stress and grain size exponents, respectively, $C$ is the material constant, $\dot{\varepsilon}$ is the secondary creep rate, $G$ is the shear modulus, σ is the applied stress, $d$ is the mean grain size, and $Q$ is the creep activation energy. The variables $Q$, $m$, and $b$ are important in differentiating various creep deformation mechanisms (see more details in [46]). For example, if $Q$ is equal to the activation energy for the solute diffusion with $m=3$ and $b=0$, this would indicates that the glide-controlled dislocation creep would be the creep mechanism [75,76]. Figure 7a shows an Arrhenius plot for the steady-state creep rates of nanocrystalline magnesium with different grain sizes for a temperature range of 573 K to 723 K. The increase in the creep rate with increasing temperature indicates that the deformation is thermally activated, indicative of a diffusion-driven creep mechanism.

The creep activation energy for 5nm and 15nm grain sizes was found to be 1.047 eV (101 kJ/mol) and 1.533 eV (148 kJ/mol), respectively. In the case of coarse-grained magnesium, the experimental range of 106 to 220 kJ/mol has been reported in the literature [24] for different applied stresses and temperature ranges. A double logarithmic plot of the steady state creep rate as a function of the applied stress for 5 nm Mg at various temperatures is shown in Figure 7b. The stress exponent, $m$, was found to be 2.67 at 673 K and about 1.77 at 723 K. Furthermore, Figure 7c shows the grain size dependence of the secondary creep rate for a temperature range of 573 K to 723 K at an applied stress of 500 MPa. At a higher temperature, the strain rate was inversely proportional to the grain size with exponent $b = $ ~1.0 – 2.0, which is comparable to the experimental work of del Valle and Ruano [76] on ultra-fine grain magnesium produced using non-equilibrium processes.

## D. Diffusional analysis with varying concentrations of Yttrium at 500 MPa

Figure 8 shows strain as a function of time for a 10 nm-grain nanocrystalline sample with different concentrations of yttrium at 673 K and with an applied constant stress of 500 MPa. 10 nm-grain sample was utilized so that all three regions of creep behavior could be observed and the role of yttrium could be assessed. Initially, for the pure nanocrystalline magnesium sample, the material is seen to deform elastically until t = 20 ps. Thereafter, the strain increases linearly at a constant creep rate, as shown in Figure 8. In the tertiary creep region (Region III), there is a sudden increase in the strain rate which results in an intergranular fracture as shown in Figure 9a. Note that similar intergranular fracture was observed in pure 3D nanocrystalline Mg specimens with varying grain sizes and morphologies, see [77]. With the addition of yttrium, the creep deformation mechanism is changed from grain boundary decohesion to grain shrinkage and rotation as shown in Figure 9. Interestingly, with the addition of yttrium, the transition from the nucleation/propagation of cracks to the enhanced grain boundary behavior was also experimentally observed under dynamic strain rate conditions in the Mg-Y alloy [19]. Creep rate for the tertiary creep region with 1 at.% Y is lower than that for 2 at.% and 3 at.% Y, as seen in Figures 8 and 10. Finally, Figure 10 shows an Arrhenius plot for the creep rates of nanocrystalline magnesium with different concentrations of yttrium for a range of temperatures from 573 K to 723 K. We therefore observe a negligible change (within the fitting error) in the overall secondary creep rate with creep activation energy changes from 1.128 to 1.154 eV for 0 to 3 at.% Y as shown in Figure 10. This could be due to a multitude of factors. For example, in present work the concentration of yttrium is limited to the GB only, which will result in significantly lower global concentrations as compared to experiments. Nevertheless, from table 3, we can conclude that the addition of yttrium increases the activation energy for grain boundary magnesium atoms compared to the activation energy of grain boundary magnesium atoms with no yttrium. Hence, the mobility of magnesium decreases with the addition of yttrium, which prevents the grain boundary decohesion (figure 9). Moreover, the addition of yttrium stabilizes the grain boundary and decreases the sliding energy of grain boundary (figure 11).

## E. Effect of yttrium on the grain boundary sliding

In the 10 nm grain size magnesium (including non-equilibrium grain boundaries) with an addition of yttrium, there was grain shrinkage and rotation as shown in Figures 9b-c. To further investigate the possibility of enhanced grain boundary sliding with the addition of yttrium, we examined the grain boundary sliding excess energy as a function of applied displacement along the <1$\bar{2}$10> direction of (10$\bar{1}$1), (10$\bar{1}$2), (10$\bar{1}$3), and (10$\bar{1}$6) boundaries using MS. Figure 11 shows the excess energy required to induce grain boundary sliding in (10$\bar{1}$1) and (10$\bar{1}$3) boundaries. Overall, the excess energy for the grain boundary sliding decreases with the addition of yttrium as compared to the pristine boundaries. On average, the decrease is about 2.7, 11.3, and 15.3 % with addition of 12.5 at.% Y for the (10$\bar{1}$2), (10$\bar{1}$3), and (10$\bar{1}$1) boundaries, respectively (Table 4). Evidence for the occurrence of enhanced grain-boundary shearing with addition of yttrium suggests that the diffusion creep must be accompanied by grain-boundary sliding. Moreover, yttrium has a softening effect on the sliding of boundaries which suggests that the experimentally observed high temperature beneficial effects of yttrium addition could be attributed to some other mechanisms or phenomena such as formation of stable yttrium oxides at grain boundaries [43,44,78], increased forest dislocation based hardening [11,20,21,42], or large grain boundary yttria and precipitate denuded zones near the grain boundaries causing material softening. Furthermore, it has been experimentally shown that in aged Mg-Y alloys, due to a supersaturated solid-solution state within the grains, dynamic precipitation (two pseudo-equilibrium phases) can occur during creep [18,26,27]. Hence, further experimental and modeling efforts are needed to elucidate the exact mechanism associated with the excellent reported creep resistance of Mg–Y solid-solution alloys.

## IV Conclusions

With the addition of yttrium, experimentally it has been shown that there is significant improvement in ductility, creep resistance, retardation of grain coarsening, grain boundary rotation/sliding [19], and corrosion resistances for coarse grained Mg-alloys. The improvement in mechanical behavior reported to be due to some combination of the following phenomena: the activation of <c+a> dislocations [11,20,21,42], forest dislocation hardening [18], inhabitance of the static recrystallization [14], and formation of an enhanced oxide film [43,44]. In this paper, we reveal and elucidate the high-temperature deformation behavior and mechanisms of nanocrystalline magnesium with and without yttrium for a temperature range of 573 K to 723 K under constant stress loadings. Our simulations show that with the addition of alloying elements such as yttrium, the creep rate in the secondary region changes negligibly. However, with the addition of yttrium, the creep deformation mechanism is changed from grain boundary decohesion to grain shrinkage and rotation. The following conclusions can be drawn from this work:

1. For nanocrystalline magnesium, in the absence of external load, we observed the transition temperature from the GB-dominant to lattice-dominant diffusion mechanism at 573 K. Similarly, in the absence of external load for the case of Mg-1 at.% Y, at a lower temperature range, the activation energy of yttrium diffusion, 0.094 eV, is lower than that of Mg, 0.279 eV. In contrast, for a higher temperature range, the activation energy of yttrium diffusion, 0.93 eV, is higher than that of magnesium, 0.659 eV. However, in the case of Mg-3 at.% Y, for high temperature, the activation energy for diffusion of magnesium and yttrium decreases when compared with Mg-1 at.% Y.
2. The creep activation for 5 nm and 15 nm grain sizes in pure nanocrystalline magnesium samples was found to be 1.047 eV (101 kJ/mol) and 1.533 eV (148 kJ/mol), respectively. In the case of coarse-grained magnesium, the experimental range of 106 to 220 kJ/mol is reported in literature [24] for different applied stresses and temperature ranges.
3. The stress exponents, $m$, were found to be 2.67 at 673 K and about 1.77 at 723 K. Furthermore, at a higher temperature, the strain rate was inversely proportional to the grain size with exponent $b = \sim 1.0 - 2.0$ which is analogous to the experimental work of del Valle and Ruano [76] using ultra-fine grain magnesium.
4. With the addition of alloying elements such as yttrium, the creep rate in the secondary region decreases and the creep deformation mechanism is changed from grain boundary decohesion to grain boundary rotation/sliding. Similarly, a change in mechanisms with addition of yttrium was observed under dynamic high strain rate experiments [19].
5. Finally, $(10\bar{1}1)$, $(10\bar{1}2)$, $(10\bar{1}3)$, and $(10\bar{1}6)$ boundary sliding simulations using MS show there is a reduction in the excess energy needed for sliding with the addition of yttrium. Hence, yttrium has a softening effect on the sliding of boundaries which suggests that the experimentally observed high temperature beneficial effects of yttrium addition is likely to be attributed to some combination of other reported creep strengthening mechanisms or phenomena such as formation of stable yttrium oxides at grain boundaries [43,44,78], increased forest dislocation based hardening [11,20,21,42], or large grain boundary yttria and precipitate denuded zones near the grain boundaries causing material softening.

Further experimental and modeling investigations are needed to elucidate the exact mechanism associated with the excellent creep resistance of Mg−Y solid-solution alloys. However, the perspectives presented here provide a physical basis through mechanism-oriented atomistic study for understanding the relationship between critical microstructure such as GB/solute interactions, and mechanical properties [79–85]. This will, in turn, provide a better understanding of critical microstructures and their role on mechanical behavior for materials design of complex structural alloys.

## Acknowledgement

The authors are grateful for the financial support for this work from the National Science Foundation award number 1463656. We also appreciate Fulton High Performance Computing at Arizona State University for enabling us to conduct our simulations.

# References

[1] Friedrich H, Schumann S. J Mater Process Technol 2001;117:276.
[2] Kainer KU, Kaiser F. 2003.
[3] Kassner M, Hayes T. Int J Plast 2003;19:1715.
[4] Luo AA. Mater Sci Forum 2003;419-422:57.
[5] Mathaudhu SN, Nyberg EA. Magnesium Alloys in Us Military Applications: Past, Current and Future Solutions. Pacific Northwest National Laboratory (PNNL), Richland, WA (US); 2010.
[6] Pollock TM. Science 2010;328:986.
[7] Watanabe H, Tsutsui H, Mukai T, Kohzu M, Tanabe S, Higashi K. Int J Plast 2001;17:387.
[8] Christian JW, Mahajan S. Prog Mater Sci 1995;39:1.
[9] Cottrell A, Bilby B. Philos Mag 1951;42:573.
[10] Yoo M. Metall Trans A 1981;12:409.
[11] Yoo MH, Morris JR, Ho KM, Agnew SR. Metall Mater Trans A 2002;33:813.
[12] Koike J, Kobayashi T, Mukai T, Watanabe H, Suzuki M, Maruyama K, Higashi K. Acta Mater 2003;51:2055.
[13] Agnew SR, Yoo MH, Tomé CN. Acta Mater 2001;49:4277.
[14] Farzadfar SA, Martin É, Sanjari M, Essadiqi E, Yue S. J Mater Sci 2012;47:5488.
[15] He SM, Zeng XQ, Peng LM, Gao X, Nie JF, Ding WJ. J Alloys Compd 2007;427:316.
[16] Kashefi N, Mahmudi R. Mater Des 2012;39:200.
[17] Lee Y, Dahle A, StJohn D. Metall Mater Trans A 2000;31:2895.
[18] Maruyama K, Suzuki M, Sato H. Metall Mater Trans A 2002;33:875.
[19] Nagao M, Terada T, Somekawa H, Singh A, Mukai T. JOM 2014;66:305.
[20] Sandlöbes S, Friák M, Zaefferer S, Dick A, Yi S, Letzig D, Pei Z, Zhu L-F, Neugebauer J, Raabe D. Acta Mater 2012;60:3011.
[21] Sandlöbes S, Zaefferer S, Schestakow I, Yi S, Gonzalez-Martinez R. Acta Mater 2011;59:429.
[22] Suzuki A, Saddock ND, Jones JW, Pollock TM. Scr Mater 2004;51:1005.
[23] Xiong Y, Yu Q, Jiang Y. Int J Plast 2014;53:107.
[24] Saddock ND. Microstructure and Creep Behavior of Mg-Al Alloys Containing Alkaline and Rare Earth Additions. University of Michigan, 2008.
[25] Bohlen J, Nürnberg MR, Senn JW, Letzig D, Agnew SR. Acta Mater 2007;55:2101.
[26] Suzuki M, Sato H, Maruyama K, Oikawa H. Mater Sci Eng A 1998;252:248.
[27] Suzuki M, Sato H, Maruyama K, Oikawa H. Mater Sci Eng A 2001;319–321:751.
[28] Chua BW, Lu L, Lai MO. Philos Mag 2006;86:2919.
[29] Inoue A, Kawamura Y, Matsushita M, Hayashi K, Koike J. J Mater Res 2001;16:1894.
[30] Seipp S, Wagner MF-X, Hockauf K, Schneider I, Meyer LW, Hockauf M. Int J Plast 2012;35:155.
[31] Youssef KM, Wang YB, Liao XZ, Mathaudhu SN, Kecskés LJ, Zhu YT, Koch CC. Mater Sci Eng A 2011;528:7494.
[32] Nabarro F. Proc Phys Soc 1947;59:256.
[33] Herring C. J Appl Phys 1950;21:437.
[34] Coble R. J Appl Phys 1963;34:1679.
[35] Raj R, Ashby M. Metall Trans 1971;2:1113.
[36] Dieter GE, Bacon D. Mechanical Metallurgy, vol. 3. McGraw-Hill New York; 1986.
[37] Gan M, Tomar V. Mater Sci Eng A 2010;527:7615.
[38] Hall EO. Proc Phys Soc Sect B 1951;64:747.
[39] Petch N. J Iron Steel Inst 1953;174:25.
[40] Zhang Y, Gan M, Tomar V. J Nanotechnol Eng Med 2014;5:021003.
[41] Domain C. J Nucl Mater 2006;351:1.
[42] Pei Z, Zhu L-F, Friák M, Sandlöbes S, Pezold J von, Sheng HW, Race CP, Zaefferer S, Svendsen B, Raabe D, Neugebauer J. New J Phys 2013;15:043020.
[43] Yao HB, Li Y, Wee ATS. Electrochimica Acta 2003;48:4197.

[44]  Liu M, Schmutz P, Uggowitzer PJ, Song G, Atrens A. Corros Sci 2010;52:3687.
[45]  Millett PC, Desai T, Yamakov V, Wolf D. Acta Mater 2008;56:3688.
[46]  Yamakov V, Wolf D, Phillpot S, Gleiter H. Acta Mater 2002;50:61.
[47]  Detor AJ, Schuh CA. Acta Mater 2007;55:4221.
[48]  Jang S, Purohit Y, Irving D, Padgett C, Brenner D, Scattergood R. Acta Mater 2008;56:4750.
[49]  Koch CC. J Mater Sci 2007;42:1403.
[50]  Mayr S, Bedorf D. Phys Rev B 2007;76:024111.
[51]  Millett PC, Selvam RP, Bansal S, Saxena A. Acta Mater 2005;53:3671.
[52]  Schäfer J, Stukowski A, Albe K. Acta Mater 2011;59:2957.
[53]  Plimpton S. J Comput Phys 1995;117:1.
[54]  Kimizuka H, Matsubara K, Ogata S. Proc 9th Int Conf Magnes Alloys Their Appl Mg2012 2012:1023.
[55]  Slutsky LJ, Garland CW. Phys Rev 1957;107:972.
[56]  Du Q, Faber V, Gunzburger M. SIAM Rev 1999;41:637.
[57]  Olson GB, Zhang S. Ductilization of High-Strength Magnesium Alloys. 2012.
[58]  Parrinello M, Rahman A. J Appl Phys 1981;52:7182.
[59]  Bhatia MA, Solanki KN. J Appl Phys 2013;114:244309.
[60]  Rhodes NR, Tschopp MA, Solanki KN. Model Simul Mater Sci Eng 2013;21:035009.
[61]  Solanki KN, Tschopp MA, Bhatia MA, Rhodes NR. Metall Mater Trans A 2013;44:1365.
[62]  Rajagopalan M, Tschopp MA, Solanki KN. JOM 2014;66:129.
[63]  Tschopp MA, Solanki KN, Gao F, Sun X, Khaleel MA, Horstemeyer MF. Phys Rev B 2012;85:064108.
[64]  Rajagopalan M, Bhatia MA, Tschopp MA, Srolovitz DJ, Solanki KN. Acta Mater 2014;73:312.
[65]  Adlakha I, Tschopp M, Solanki K. Mater Sci Eng A 2014.
[66]  Adlakha I, Bhatia MA, Tschopp MA, Solanki KN. Philos Mag 2014;94:3445.
[67]  Adlakha I, Solanki KN. Sci Rep 2015;5.
[68]  Dai S, Xiang Y, Srolovitz DJ. Acta Mater 2013;61:1327.
[69]  Sutton AP, Vitek V. Philos Trans R Soc Lond Ser Math Phys Sci 1983;309:1.
[70]  Ratanaphan S, Olmsted DL, Bulatov VV, Holm EA, Rollett AD, Rohrer GS. Acta Mater 2015;88:346.
[71]  Wang J, Beyerlein IJ. Model Simul Mater Sci Eng 2012;20:024002.
[72]  Mantina M, Chen LQ, Liu ZK. Predicting Diffusion Coefficients from First Principles via Eyring's Reaction Rate Theory, in:. Defect Diffus. Forum, vol. 294. Trans Tech Publ; 2009.
[73]  Johnson WA. Transactions 1942:331.
[74]  Fisher JC. J Appl Phys 1951;22:74.
[75]  Sherby OD, Wadsworth J. Prog Mater Sci 1989;33:169.
[76]  Del Valle JA, Ruano OA. Acta Mater 2007;55:455.
[77]  Moitra A. Comput Mater Sci 2013;79:247.
[78]  Han BQ, Dunand DC. Mater Sci Eng A 2000;277:297.
[79]  Ahad F, Enakoutsa K, Solanki K, Bammann D. Int J Plast 2014;55:108.
[80]  Bammann DJ, Solanki KN. Int J Plast 2010;26:775.
[81]  Fernández A, Pérez Prado MT, Wei Y, Jérusalem A. Int J Plast 2011;27:1739.
[82]  Horstemeyer MF, Bammann DJ. Int J Plast 2010;26:1310.
[83]  Li D, Ahzi S, M'Guil S, Wen W, Lavender C, Khaleel M. Int J Plast 2014;52:77.
[84]  Wang H, Wu P, Wang J. Int J Plast 2013;47:49.
[85]  Solanki KN, Bammann DJ. Acta Mech 2010;213:27.